\documentclass[journal,transmag]{IEEEtran}
\usepackage[rgb]{xcolor}
\usepackage{makecell}
\usepackage{graphicx}
\graphicspath{{figures/}}
\newif\ifmanuscriptbibliographyavailable
\IfFileExists{references.bib}
  {\manuscriptbibliographyavailabletrue}
  {\manuscriptbibliographyavailablefalse}
\usepackage{amsmath,amssymb,amsfonts}
\usepackage{algorithmic}
\usepackage{algorithm}
\usepackage{array}
\usepackage{textcomp}
\usepackage{stfloats}
\usepackage{url}
\usepackage{verbatim}
\usepackage{cite}
\usepackage{comment}
\usepackage{multirow}
\usepackage{booktabs}
\usepackage{bm}
\usepackage{float}
\usepackage{tabularx}
\newif\ifreviewlinenumbers
\reviewlinenumbersfalse
\ifreviewlinenumbers
  \usepackage[switch]{lineno}
  \newcommand*\patchAmsMathEnvironmentForLineno[1]{%
    \expandafter\let\csname old#1\expandafter\endcsname\csname #1\endcsname
    \expandafter\let\csname oldend#1\expandafter\endcsname\csname end#1\endcsname
    \renewenvironment{#1}%
      {\linenomath\csname old#1\endcsname}%
      {\csname oldend#1\endcsname\endlinenomath}}%
  \newcommand*\patchBothAmsMathEnvironmentsForLineno[1]{%
    \patchAmsMathEnvironmentForLineno{#1}%
    \patchAmsMathEnvironmentForLineno{#1*}}%
  \AtBeginDocument{%
    \patchBothAmsMathEnvironmentsForLineno{equation}%
    \patchBothAmsMathEnvironmentsForLineno{align}%
    \patchBothAmsMathEnvironmentsForLineno{gather}}%
\fi

\begin{document}
\ifmanuscriptbibliographyavailable
  \bstctlcite{IEEERepeatedAuthorControl}
\fi
\flushbottom

\title{Methods for Accelerating and Stabilizing Nonlinear Magnetic Equivalent-Circuit Analysis}

\author{\IEEEauthorblockN{Danial Kazemikia,
Matthew C. Gardner,~\IEEEmembership{Member,~IEEE}}
\IEEEauthorblockA{Department of Electrical Engineering, University of Texas at Dallas, Richardson, Texas}%
\thanks{{\raggedright Corresponding author: D. Kazemikia (email: danialkazemikia@utdallas.edu).\par}}}

\IEEEtitleabstractindextext{%
\begin{abstract}
Design and optimization of electromagnetic devices require models that converge reliably and quickly and can represent intense localized magnetic saturation. This paper presents a parameterized 2D nonlinear magnetic equivalent circuit (MEC) and proposes solution methods for it: an approximate Jacobian that is symmetric positive definite, so each Newton--Raphson iteration can use sparse Cholesky factorization with fill-reducing reordering, and Anderson mixing as step control. These methods are compared with alternative nonlinear iterations, linear solvers, and step-control methods in convergence, computational cost, and prediction accuracy. Radial-flux magnetic gears (RFMGs) with structural bridges serve as a challenging test case for these methods, because their thin bridges exhibit intense local saturation. To isolate the effect of this saturation, the methods are evaluated on three bridged base designs and on 1,024 bridged designs, each paired with the same design but without the bridge. On the three base designs, Newton--Raphson iteration with the approximate Jacobian reached much lower residuals than fixed-point iteration, which converged for none of them, with or without relaxation. Sparse Cholesky factorization with reordering was the fastest linear solver tested. Across the 1,024 design pairs, full-step Newton--Raphson converged for 193 bridged and 564 bridge-free designs, whereas Anderson mixing converged for 1,003 and 1,022 and required fewer iterations. Under every method, bridged designs converged less often and more slowly than their bridge-free counterparts. For the bridged and bridge-free populations, respectively, the nonlinear MEC was 2.17$\times$ and 1.84$\times$ faster than fixed-mesh nonlinear finite element analysis (FEA) in total computation time. The median outer-rotor torque discrepancy relative to FEA was 0.22\% across all designs, and the Spearman rank correlations of the design rankings were at least 0.9996.
\end{abstract}

\begin{IEEEkeywords}
Anderson mixing, approximate Jacobian, Cholesky factorization, fill-reducing reordering, finite element analysis, fixed-point iteration, inexact Newton method, line search, magnetic equivalent circuit, magnetic saturation, modulator bridge, radial-flux magnetic gear, reluctance network
\end{IEEEkeywords}}

\maketitle

\IEEEdisplaynontitleabstractindextext

\IEEEpeerreviewmaketitle
\ifreviewlinenumbers\linenumbers\fi

\section{Introduction}
\label{sec:introduction}
\IEEEPARstart{E}{lectromagnetic} models trade accuracy against computational cost depending on the physical effects they represent. Subdomain models calculate the air-gap field quickly, but treat the iron as infinitely permeable \cite{Lubin2010Analytical,Akcay2020Analytical} or linear \cite{Desvaux2017Computation,Desvaux2019Development}, so they cannot represent saturation. Their error increases with saturation \cite{Zhao2021Aa}, limiting aggressive optimization \cite{Desvaux2019Development}. Harmonic models that include the nonlinear B--H curve remove this limitation, but can take about 80\% of the finite element analysis (FEA) computation time \cite{Zhao2021Aa}, providing little advantage. Subdomain and harmonic models are also tailored to specific geometries. They can handle changes in dimensions, but their equations must be derived again when the shape of any component changes. High-resolution magnetic equivalent circuits (MECs), also known as reluctance networks, offer a middle ground between these methods. They represent nonlinear iron through the permeability of each flux tube, and reluctance-network analysis of a magnetic gear has been reported to take seconds where FEA took about 30~minutes \cite{Fukuoka2011Dynamic}. Other MECs have been reported to run 12 to 271$\times$ faster than FEA \cite{Johnson2018Parameterizeda,Amrhein20093,Korman2025Distributed}. However, speedups of MEC implementations from different studies cannot be compared directly, because each was measured against FEA with different settings rather than against a common baseline. The FEA time depends on the mesh density \cite{Jansson2020Convergence} and the adaptive refinement procedure \cite{Yamazaki2003Adaptive}, and the largest speedup above was measured against FEA with a 0.1\% energy-error target and at least eight adaptive passes, which its authors identified as part of the speedup \cite{Johnson2018Parameterizeda}. To allow a direct comparison, this paper solves the same designs with the nonlinear MEC and a common FEA baseline, which uses a fixed mesh and the same convergence criterion (Subsection~\ref{subsec:fea_reference}).

Because MECs are fast, they are attractive for population and optimization studies of hundreds or thousands of designs. Such studies need a parameterized model, which constructs the geometry, material assignments, and reluctance network from the design variables and applies the same discretization, initialization, solver, and convergence procedures to every design (Subsections~\ref{subsec:meshing}, \ref{subsec:newton_raphson}, and~\ref{subsec:convergence_criterion}). A nonparameterized MEC, in contrast, is adjusted manually for each case, which is impractical at this scale and makes its accuracy depend on those adjustments. A parameterized model avoids both problems by using one solution procedure that converges reliably and quickly for every design.

How reliably and quickly the solution converges across designs depends on how the nonlinear MEC is formulated and solved. Several earlier nonlinear MECs have been solved with Newton--Raphson iteration \cite{Derbas2009A,Diab2023nonlinear,Desenfans2024Comparison}, which linearizes the equations at the current state and solves a linear system with the Jacobian at each iteration. The Jacobian contains derivative terms of the reluctances with respect to the fluxes, and forming these terms has been identified as the cost of using Newton--Raphson \cite{Diab2023nonlinear}. The derivative terms depend on which fluxes set the permeability of each element. In some earlier MECs, flux tubes follow flux paths chosen in advance \cite{Amrhein20093,Derbas2009A,Boomiraja2020Convergence}, so the permeability of each tube depends only on that tube's own flux, and each derivative term involves a single tube \cite{Derbas2009A}. However, predetermined flux paths reduce the flexibility of the model, because paths chosen for one geometry may not hold when the design changes. Some distributed MECs \cite{Korman2025Distributed} keep this single-tube structure by neglecting the flux in the other direction, at the cost of judging saturation from one flux component rather than from the flux-density vector. In contrast, our mesh does not require determining the flux direction in advance, so the permeability of a cell depends on both its radial and tangential fluxes (Subsection~\ref{subsec:meshing}), and its derivative terms couple the branches that meet in the cell. These coupling terms make the exact Jacobian nonsymmetric, so Cholesky factorization, which requires a symmetric positive-definite matrix, cannot solve the update.

An approximate Jacobian that omits these coupling terms avoids forming them, and it is symmetric positive definite, so Cholesky factorization can solve each update (Subsection~\ref{subsec:approx_jacobian}). Newton--Raphson iteration that uses such an approximation in place of the exact Jacobian is an inexact Newton iteration. To our knowledge, the only approximate Jacobians applied to MECs before are the apparent-permeance matrix, used in a node-MMF model where the exact Jacobian failed to converge \cite{Derbas2009A}, and a quasi-Newton update of an exact Jacobian \cite{Desenfans2024Comparison}. Neither is derived from the exact Jacobian by identifying the terms it omits, so neither shows when the approximation is exact.

Because the approximate Jacobian omits the coupling terms, the inexact Newton iteration may lose its advantage over fixed-point iteration, which avoids the Jacobian altogether. This paper therefore tests whether the inexact Newton iteration still converges better than fixed-point iteration.

Besides forming the Jacobian, each Newton--Raphson iteration must solve a linear system with it, and the cost of this solve depends on how the matrix is stored, which method solves it, and whether the matrix is reordered and scaled before the solve. MEC matrices are commonly stored in sparse form, but the time and memory saved over a dense solution were not measured \cite{Johnson2018Parameterized,Diab2023nonlinear}. The incomplete-Cholesky conjugate gradient method (ICCG) and Gaussian elimination have been used in nonlinear magnetic FEA \cite{Nakata1992Improvements,Fujiwara2005The}, and Wojciechowski et al. used Cholesky factorization within block relaxation to solve a linear system of coupled reluctance--resistance network equations \cite{Wojciechowski2010Inducted}. To our knowledge, no previous MEC study has reported using Cholesky factorization for nonlinear magnetic equations or evaluated fill-reducing reordering or Jacobi scaling for nonlinear MEC analysis.

Fast linear solves are useful only if the iteration also converges reliably. Step control aims to make Newton--Raphson iteration converge more reliably by limiting or combining its updates. Step control matters for MECs because saturation reduces the differential permeability by orders of magnitude, so a full correction can overshoot. Nonlinear magnetic FEA studies reported oscillation or nonconvergence with full-step Newton--Raphson iterations \cite{Nakata1992Improvements,Neagoe1994Analysis} and investigated relaxation and line search as remedies \cite{Fujiwara2005The,Takahashi2007Large}, and in a magnetic-gear MEC, Diab et al. damped the Newton--Raphson update with a factor chosen at each iteration to reduce the residual norm \cite{Diab2023nonlinear}. Another approach is Anderson mixing, which accelerates an iterative method by combining its current and previous iterates. It is widely used in electronic structure calculations \cite{Walker2011Anderson} and has been applied to nonlinear electromagnetic FEA \cite{Filippini2019Anderson}. Applied to Newton--Raphson iteration, it acts as step control, because the combined state replaces the plain Newton--Raphson update. To our knowledge, no previous MEC study has compared step-control methods across a population of designs or applied Anderson mixing.

\begin{figure}[!t]
\centering
\includegraphics{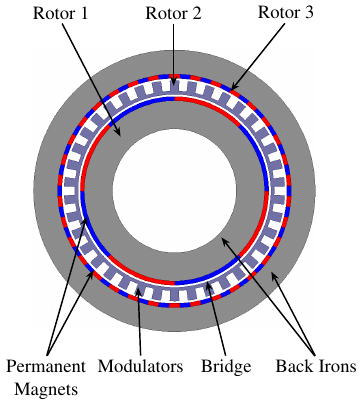}
\caption{Components of a coaxial radial-flux magnetic gear with surface-mounted PMs.}
\label{RFMG}
\end{figure}

The methods must also converge for heavily saturated designs, and radial-flux coaxial magnetic gears (RFMGs) with bridges provide a challenging test case for this. Magnetic gears transfer torque between a high-speed, low-torque shaft and a low-speed, high-torque shaft through modulated magnetic fields rather than through contacting mechanical teeth. Their contactless operation can provide overload protection, physical isolation between the shafts, and lower maintenance and acoustic noise than conventional geared drivetrains \cite{Tao2024Experimental}, which makes them attractive for wind- and wave-energy systems, electrified-aircraft propulsion, and space robotics \cite{Frank2009Gearing,Uppalapati2012Performance,Scheidler2018NASA,Scheidler2023Detailed,Praslicka2021Practical}. Despite these advantages, magnetic gears remain less mature than many electrical machines and mechanical geartrains, and most of their analyses have relied on computationally expensive FEA \cite{Praslicka2022Review,Tao2024Experimental,Frank2009Gearing,Uppalapati2012Performance,Frandsen2015Motor,Scheidler2018NASA,atallah2001,Scheidler2023Detailed,Praslicka2021Practical}. RFMGs have achieved the highest experimentally demonstrated torque densities among magnetic-gear topologies \cite{Praslicka2022Review,Wong2023Experimentally}. As illustrated in Fig. \ref{RFMG}, an RFMG has an inner rotor (Rotor 1), a modulator rotor (Rotor 2), and an outer rotor (Rotor 3). RFMGs often include a bridge connecting the ferromagnetic modulators, which increases the stiffness of Rotor 2 and simplifies manufacturing because all modulators can be formed from a single lamination stack \cite{Modaresahmadi2021Structural,Li2019Designing,Zhan2019Torque}. The bridge forms a thin flux leakage path with low reluctance until it saturates, and its geometry can substantially change the magnetic field distribution and torque. Bridged RFMGs therefore test whether a nonlinear MEC solution method handles severe localized saturation consistently, rapidly, and accurately. They also test whether the nonlinear solution is needed at all, because a linear MEC, which cannot represent saturation, may suffice for the same designs without the bridge. To our knowledge, how bridges affect the accuracy and convergence of a nonlinear MEC has not been examined before.

In summary, this paper makes four contributions to the solution of nonlinear MECs. The first contribution is an approximate Jacobian for cells whose permeability depends on both flux components, derived from the exact Jacobian by identifying and omitting the terms that couple the branches meeting in a cell. It is symmetric positive definite, so sparse Cholesky factorization can solve each update, and for single flux tubes it reduces to the exact Jacobian in \cite{Boomiraja2020Convergence}. The resulting inexact Newton iteration is compared with fixed-point iteration to test whether omitting these terms costs it its advantage. The second contribution is an evaluation of fill-reducing reordering and Jacobi scaling for sparse Cholesky factorization and ICCG. The third contribution is Anderson mixing applied to the resulting inexact Newton iteration, compared with full-step Newton--Raphson and line search across a population of designs. The fourth contribution is an evaluation of how bridge saturation affects the convergence and accuracy of these methods, using bridged designs paired with the same designs without the bridge.

These methods are evaluated using a parameterized nonlinear MEC for three bridged RFMG base designs and 1,024 bridged designs. Each of the 1,024 is also analyzed with its bridge removed to isolate the effect of bridge saturation. The comparisons address convergence, computational cost, and prediction accuracy (Section~\ref{sec:evaluation}). The mesh-flux and node-MMF formulations are compared first (Subsection~\ref{subsec:formulation_comparison}), then inexact Newton and fixed-point iteration (Subsection~\ref{subsec:newton_vs_fixed_point}). For the linear solve, sparse Cholesky factorization and ICCG are first tested with and without reordering and scaling (Subsection~\ref{subsec:linear_solver_results}). Explicit inversion is then compared with sparse Cholesky factorization on meshes of increasing size (Subsection~\ref{subsubsec:dense_vs_cholesky}), and sparse Cholesky factorization with conjugate gradient (CG), ICCG, and sparse lower--upper (LU) factorization (Subsection~\ref{subsubsec:cholesky_vs_alternatives}). Full-step Newton--Raphson, line search, and Anderson mixing are compared across the design population (Subsection~\ref{subsec:convergence}). Nonlinear FEA provides the reference for prediction accuracy, design ranking, and computation time (Subsections~\ref{subsec:fea_reference} to~\ref{subsec:population_accuracy}). The paired designs show how bridges affect convergence difficulty, saturation severity, and linear-MEC accuracy (Subsections~\ref{subsec:convergence}, \ref{subsec:population_accuracy}, and~\ref{subsec:bridge_effects}).

\section{MEC Formulation}
\label{sec:methods}

\subsection{Meshing and the Node Cell}
\label{subsec:meshing}
\begin{figure}[!t]
\centering
\includegraphics[trim=15pt 8pt 15pt 8pt,clip,width=\columnwidth]{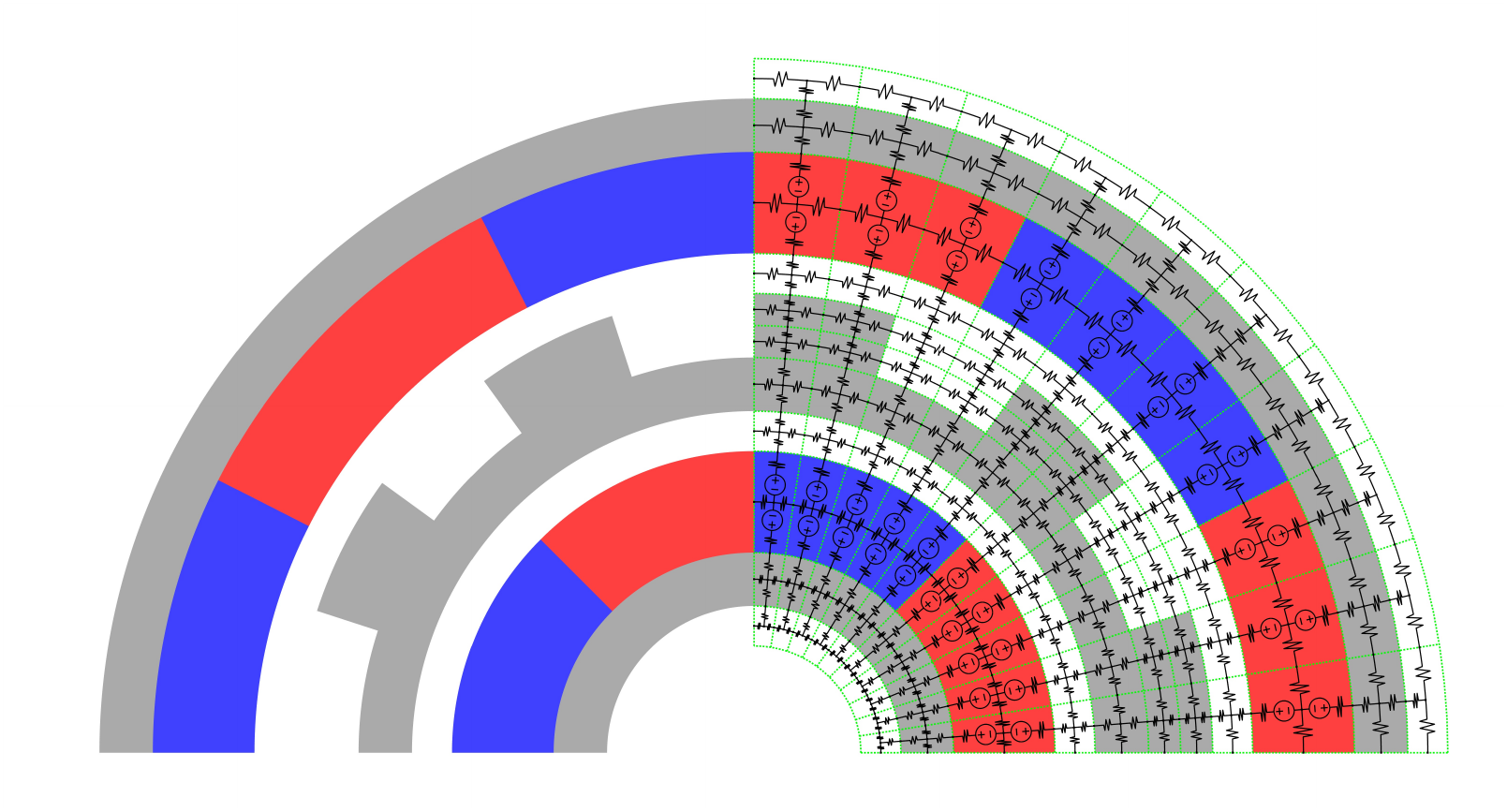}
\caption{RFMG semicircle showing a coarse MEC network (0°–90°) and the material layout (90°–180°).}
\label{Network Overlay}
\end{figure}

Any two-dimensional flux vector can be resolved into perpendicular radial and tangential components. The mesh therefore provides paths in both directions without requiring the flux direction in advance. The two-dimensional cross-section is discretized by radial and angular boundaries, whose intersections define arc-shaped node cells (dotted green lines in Fig.~\ref{Network Overlay}). Each node cell contains four half-branches, two radial and two tangential, which carry the two flux components and share the cell’s permeability. Each half-branch represents a flux tube connecting the cell center to one of its four faces and is assigned a reluctance, following \cite{Johnson2018Parameterized}. Two half-branches meeting at a shared face form a branch connecting adjacent cell centers and carrying a single flux.

\begin{figure}[!t]
\centering
\includegraphics{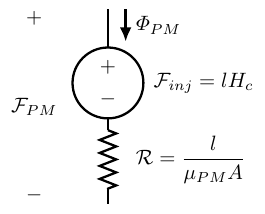}
\caption{Equivalent circuit of a radial PM flux tube with length $l$, magnet coercivity $H_c$, PM permeability $\mu_{PM}$, and cross-sectional area $A$.}
\label{equivalent circuit PM}
\end{figure}

Each permanent-magnet (PM) flux tube is represented by a Thevenin equivalent circuit consisting of a reluctance in series with a magnetomotive force (MMF) source (Fig.~\ref{equivalent circuit PM}). All PMs here are radially magnetized, so only radial flux tubes contain MMF sources. Halbach and spoke-type arrangements would also require tangential MMF sources.

The model partitions the inner back iron, inner PMs, inner air gaps, bridge, modulators, outer air gaps, outer PMs, and outer back iron into a grid of node cells. Inner bore and exterior air regions extend the numerical domain beyond the physical back irons.

For each PM, air gap, bridge, and modulator region, the radial layer count is $N_{RL}=\max\left(\left\lceil 20T/\ell_p\right\rceil,3\right)$, where $T$ is the regional radial thickness, $\ell_p=\pi(R_{in}+R_{out})/(2P)$ is its local magnetic pole pitch, $R_{in}$ and $R_{out}$ are the region's inner and outer radii, and $P$ is its pole-pair or modulator count for that region. This rule concentrates radial resolution where the radial field gradients are expected to be greatest. The remaining regions use two radial layers, because the fields in the inner bore and exterior air are expected to be too weak to affect torque, and the flux density in the back irons is expected to vary less radially than in other regions.

When the geometry and permanent-magnet excitation permit circumferential symmetry, MEC and FEA solve only one sector. Periodic or antiperiodic conditions represent the repeated sectors, and torque is reported for the complete gear, as in \cite{Johnson2018Parameterized}.

The angular mesh is nonuniform. Cell edges are placed on every boundary between two materials and on the edges of the symmetry sector, so that no cell contains two materials. Between these edges, cells are subdivided so that each modulator pitch ($360/Q_2$ degrees, where $Q_2$ is the total number of modulators) contains at least 60 cells. All radial bands share the same angular edges.

\subsection{Mesh-Flux Formulation and Matrix Structure}
\label{subsec:mesh_flux}

\begin{figure}[!t]
\centering
\includegraphics[scale=0.85]{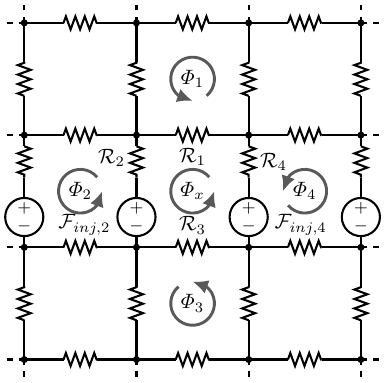}
\caption{Mesh loop $\varPhi_x$ in a PM layer and its adjacent mesh loops.}
\label{2D Mesh loop}
\end{figure}

The mesh-flux approach writes one equation for each unknown loop flux. Applying Ampere's circuital law to the loop in Fig. \ref{2D Mesh loop} yields

\begin{equation}
\label{Loop eq}
\sum_{i=1}^{4} {\mathcal{R}}_{i}(\varPhi) ({\varPhi}_{x}-{\varPhi}_{i})  = \mathcal{F}_{inj,4} - \mathcal{F}_{inj,2},
\end{equation}

The steel permeability makes the loop reluctances flux-dependent, giving $N$ nonlinear loop equations to solve for the $N$ loop fluxes:

\begin{equation}
\label{Nonlinear System eq}
[\mathcal{R}(\varPhi)]_{N \times N} [\varPhi]_{N \times 1} = [f]_{N \times 1},
\end{equation}
Here $\mathcal{R}$ is the reluctance matrix and $\varPhi$ contains the unknown mesh fluxes. The left-hand side gives the net MMF drop across the reluctances in each loop, and $f$ contains the corresponding net source MMFs, which are zero for loops without PMs.

The residual of these equations,

\begin{equation}
\label{residual eq2}
r(\varPhi)=\mathcal{R}(\varPhi)\,\varPhi - f,
\end{equation}

\noindent has the MMF imbalances of the mesh loops as its entries.

In row $i$ of $\mathcal{R}$, the diagonal entry is the sum of the reluctances around loop $i$, and its off-diagonal entries are the negative reluctances shared with the adjacent loops. An interior loop has four neighbors, so each row has at most five nonzero entries and $\mathcal{R}$ is sparse. The reluctance of a branch is the same in either direction, so $\mathcal{R}$ is also symmetric.

\subsection{Node-MMF Formulation}
\label{subsec:node_mmf}
The node-MMF formulation uses the MMFs at the cell centers as unknowns. For the branch joining cells $c$ and $d$,

\begin{equation}
\label{half branch eq}
\mathcal R_{cd}=\frac{\ell_c}{\mu_c A_c}+\frac{\ell_d}{\mu_d A_d},
\qquad
\phi_{cd}=\frac{F_c-F_d+\mathcal F_{cd}}{\mathcal R_{cd}},
\end{equation}

\noindent where $F_c$ and $F_d$ are the node MMFs, $\mathcal F_{cd}$ is the branch source MMF, and $\ell$, $A$, and $\mu$ are the half-branch length, cross-sectional area, and cell permeability. Conservation of magnetic flux at each node, $\sum_d \phi_{cd}=0$ over the branches connected to node $c$, gives one equation per node in the node MMFs. The residual is the flux imbalance at each node.

The cell permeability $\mu_c$ is evaluated from all four half-branch fluxes (Subsection~\ref{subsec:bh_material}), including the branch flux that (\ref{half branch eq}) calculates from it. This dependency is circular, so (\ref{half branch eq}) uses the permeabilities of the previous iteration. Each node is at a cell center, so each branch joins two cells, and the potentials alone do not give the field inside one cell. The field must therefore be computed from the branch fluxes. If the nodes are placed so that each cell has its own potentials, the field intensity in a cell follows from those potentials alone, and the permeability does not depend on the branch fluxes \cite{Amrhein20093}. Such formulations are not examined here.

\subsection{Nonlinear B--H Material Model}
\label{subsec:bh_material}

The nonlinear properties of M250 electrical steel are represented by a $B$--$H$ curve interpolated from measured points up to $B_k=1.95$ T using the modified Akima method \cite{Akima1970Interpolation}. Above $B_k$, the curve continues with the slope of free space, as in fully saturated iron. A linear MEC instead fixes the relative permeability of M250, so it cannot represent saturation.

\begin{equation}
H(B)=
\begin{cases}
H_{\mathrm{Akima}}(B), & 0 \leq B \leq B_k,\\
H_{\mathrm{Akima}}(B_k)+\dfrac{B-B_k}{\mu_0}, & B>B_k.
\end{cases}
\label{bh_material_law}
\end{equation}

The apparent permeability, $\mu$, is used to evaluate the nonlinear MEC residual, and the differential permeability, $dB/dH$, is used to assemble the Jacobian (Subsection~\ref{subsec:approx_jacobian}). Both are evaluated at the magnitude of the cell flux-density vector. Its radial and tangential components are calculated from the corresponding half-branch fluxes:

\begin{gather}
\label{cell flux density eq components}
B_r=\frac{\phi_{\mathrm{in}}+\phi_{\mathrm{out}}}{2A_r},
\qquad
B_\theta=\frac{\phi_{\mathrm{cw}}+\phi_{\mathrm{ccw}}}{2A_\theta}\\
\label{cell flux density eq}
B=\sqrt{B_r^2+B_\theta^2},
\end{gather}

\noindent where, $B_r$ and $B_\theta$ are the average radial and tangential flux densities in the cell. The fluxes $\phi_{\mathrm{in}}$, $\phi_{\mathrm{out}}$, $\phi_{\mathrm{cw}}$, and $\phi_{\mathrm{ccw}}$ correspond to fluxes in the radially inner, radially outer, clockwise, and counter-clockwise half-branches, respectively. $A_r$ and $A_\theta$ are the average cross-sectional areas of the cell's radial and tangential half-branches. All four half-branches of the cell share the same permeability $\mu$. At each iteration, the updated mesh fluxes give the branch fluxes, from which these quantities are reevaluated.

\subsection{Torque Calculation}
\label{subsec:torque_calculation}

The torque per unit of axial length is calculated with the Maxwell stress tensor \cite{Amrhein2009Force} in the middle cell layer of each air gap. The inner and outer air-gap layers give the torques on Rotors 1 and 3, and the Rotor 2 torque follows from torque balance. Each design is evaluated at a single rotor position, at the maximum torque angle, which can represent the slip torque accurately because the designs have noninteger gear ratios and therefore little torque ripple \cite{Praslicka2022Review} (Subsection~\ref{subsec:study_designs}).

\section{Nonlinear Solution Methods}
\label{sec:solution}

\subsection{Nonlinear Iteration Methods}
\label{subsec:newton_raphson}

\subsubsection{Newton--Raphson Method}
\label{subsubsec:newton_iteration}
Newton--Raphson iteration solves a nonlinear system $r(u)=0$ for an unknown vector $u$ by repeating

\begin{align}
\label{Newton Raphson eq}
&J(u_k)\Delta u_k=-r(u_k)\\
\label{Newton Raphson eq update}
&u_{k+1}=u_k+\alpha_k\Delta u_k,
\end{align}

\noindent where $r$ is the residual, $J$ is the Jacobian of $r$, and $\alpha_k$ is the step length applied to the correction.

\subsubsection{Fixed-Point Iteration}
\label{subsubsec:fixed_point_iteration}
Fixed-point iteration instead avoids the Jacobian by evaluating reluctances at the current mesh fluxes and solving $\mathcal{R}(\varPhi_k)\hat{\varPhi}_{k+1}=f$. The update is relaxed as $\varPhi_{k+1}=\varPhi_k+\beta(\hat{\varPhi}_{k+1}-\varPhi_k)$, where $\beta=1$ applies no relaxation.

\subsubsection{Initialization}
\label{subsec:initialization}
Both iterations need an initial state, $u_{k=0}$ in (\ref{Newton Raphson eq}) and $\varPhi_0$ in fixed-point iteration. Both start from the linear MEC solution of (\ref{Nonlinear System eq}) assuming a relative permeability of 4000 for the steel, so every comparison begins from the same state.

\subsection{Convergence Criterion}
\label{subsec:convergence_criterion}
At the exact solution, every MEC equation is satisfied and the residual is zero, giving a known target for assessing convergence. We summarize the residual vector $r(\varPhi_k)$ by its L2 norm, $\lVert r(\varPhi_k)\rVert_2$, a scalar measure of the total MMF imbalance. The residual is normalized by the source-MMF norm:

\begin{equation}
\label{residual normalized eq}
\varepsilon_k=\frac{\lVert r(\varPhi_k)\rVert_2}{\lVert f\rVert_2},
\end{equation}

\noindent where $f$ contains the MMF sources of the mesh-loop equations. The L2 norm scales with the excitation, which differs between designs. The ratio is dimensionless and measures the remaining imbalance relative to the excitation, which makes one criterion comparable across designs. Likewise, the node-MMF residual, a flux balance, is divided by the norm of its own source vector and is also dimensionless. For the remainder of this paper, ``normalized residual'' means the source-normalized residual $\varepsilon_k$. A solution is considered converged when this normalized residual is at or below $10^{-4}$ on two consecutive iterations, with at least three and at most 50 iterations. This threshold means that the residual norm is at most 0.01\% of the source norm. An iteration counts one applied nonlinear correction, and rejected trial steps do not count.

The same $10^{-4}$ threshold is applied to every design and every solution method, so all convergence results are compared on a common basis. The threshold is a small value chosen to suit any design; it was not tuned to the designs in this paper or to agreement with FEA torque. How the results change with other thresholds, or with other stopping criteria such as update size or changes in torque or flux density, is beyond the scope of this paper and is left for future work.

\subsection{Approximate Jacobian and Inexact Newton Iteration}
\label{subsec:approx_jacobian}
The Newton--Raphson iteration uses an approximation $\widetilde J$ of the Jacobian $J$ in place of $J$, which makes it an inexact Newton iteration. Because each ferromagnetic cell has one permeability, set by both its radial and tangential flux densities (Subsection~\ref{subsec:bh_material}), a change in either component changes the reluctances in both directions. These inter-branch dependencies make the exact Jacobian nonsymmetric. A symmetric positive-definite approximation is derived below to enable sparse Cholesky factorization.

For mesh loop $x$ in Fig.~\ref{2D Mesh loop}, the residual $r_x$ is the left-hand side of (\ref{Loop eq}) minus its right-hand side $f_x$, the net source MMF in the loop. Differentiating with respect to mesh flux $\varPhi_j$ gives the exact Jacobian entry,
\begin{equation}
\label{eq:jacobian_exact_entry}
J_{xj}=\sum_{i=1}^{4}\bigg[\mathcal R_i\frac{\partial(\varPhi_x-\varPhi_i)}{\partial\varPhi_j}
+(\varPhi_x-\varPhi_i)\frac{\partial\mathcal R_i}{\partial\varPhi_j}\bigg].
\end{equation}
The first term describes how the MMF drop changes with branch flux at fixed reluctance. The second accounts for the change in reluctance as the flux changes the cell permeability.

For half-branch $h$, $\mathcal F_h=(\ell_h/A_h)\,\nu(B)\,\phi_h$ gives the MMF drop, where $\ell_h$ and $A_h$ are the length and area of its flux tube and $\nu$ is the reluctivity of its cell, evaluated from the cell flux density $B$ in (\ref{cell flux density eq}). Its exact derivative is
\begin{equation}
\label{eq:jacobian_exact_branch_response}
\frac{\partial\mathcal F_h}{\partial\phi_k}
=\mathcal R_h\frac{\partial\phi_h}{\partial\phi_k}+\phi_h\frac{\partial\mathcal R_h}{\partial\phi_k}.
\end{equation}
By (\ref{cell flux density eq components}) and (\ref{cell flux density eq}), $B$ depends on the fluxes of all four half-branches of the cell, with $\partial B/\partial\phi_k=B_{d(k)}/(2A_{d(k)}B)$, where $B_{d(k)}$ is the component $B_r$ or $B_\theta$ along half-branch $k$ and $A_{d(k)}$ is the corresponding average path area. For a different half-branch $k$ of the same cell, the derivative is therefore
\begin{equation}
\label{eq:jacobian_cross_branch}
\frac{\partial\mathcal F_h}{\partial\phi_k}
=\frac{\ell_h}{A_h}\,\phi_h\,\frac{d\nu}{dB}\,\frac{B_{d(k)}}{2A_{d(k)}B},\qquad k\ne h.
\end{equation}
Exchanging $h$ and $k$ gives the same value only if
\begin{equation}
\label{jacobian symmetry condition eq}
\frac{\ell_h\phi_hB_{d(k)}}{A_hA_{d(k)}}=\frac{\ell_k\phi_kB_{d(h)}}{A_kA_{d(h)}}.
\end{equation}
\noindent This does not hold in general, because the half-branches of a cell carry different fluxes and have different lengths and areas. The exact Jacobian is therefore not symmetric, and Cholesky factorization cannot be applied to it.

To obtain a symmetric matrix, the approximate Jacobian omits these inter-branch terms, so that each half-branch responds only to its own flux,
\begin{equation}
\label{eq:jacobian_approx_branch_response}
\frac{\partial\mathcal F_h}{\partial\phi_k}
\approx\widetilde{\mathcal R}_h\frac{\partial\phi_h}{\partial\phi_k}.
\end{equation}
The correction reluctance $\widetilde{\mathcal R}_h$ follows from the material tangent of the cell. For isotropic steel at $\lVert\mathbf B\rVert>0$, the inverse material tangent is
\begin{equation}
\label{eq:inverse_tangent_reluctivity}
\frac{\partial\mathbf H}{\partial\mathbf B}=\nu(I-\mathbf u\mathbf u^T)+\nu_d\mathbf u\mathbf u^T.
\end{equation}
where $\mathbf u=\mathbf B/\lVert\mathbf B\rVert$ and $\nu_d=d\lVert\mathbf H\rVert/d\lVert\mathbf B\rVert=\nu+\lVert\mathbf B\rVert\,d\nu/d\lVert\mathbf B\rVert$ is the differential reluctivity.
Because $\widetilde{\mathcal R}_h$ acts on one half-branch alone, its reluctivity must be one value per cell, shared by both directions, as the apparent reluctivity is. For a flux change along the field direction,
\begin{equation}
\label{eq:tangent_exact_along_field}
\frac{\partial\mathbf H}{\partial\mathbf B}\,\mathbf u=\nu_d\mathbf u,
\end{equation}
so $\nu_d$ is the only such value that is exact for a flux change along the field direction. The correction reluctance of half-branch $h$ is therefore $\widetilde{\mathcal R}_h=(\nu_d\ell_h/A_h)$, which is the reluctance of the half-branch calculated using the differential reluctivity, rather than the apparent reluctivity. Material properties are evaluated at the current state.

The omitted part of the inverse tangent is
\begin{equation}
\label{eq:tangent_omitted}
\frac{\partial\mathbf H}{\partial\mathbf B}-\nu_dI=(\nu-\nu_d)(I-\mathbf u\mathbf u^T).
\end{equation}
The omitted material-tangent term vanishes for a linear material. For a nonlinear material, its projection along a half-branch vanishes when the field is parallel to that half-branch. For an isolated flux tube with uniform cross-sectional area, the flux density is $B=\phi/A$, so differentiating its MMF drop gives
\begin{equation}
\label{eq:single_tube_differential_reluctance}
\frac{d\mathcal F}{d\phi}
=\frac{\ell}{A}\frac{dH}{dB}
=\frac{\ell}{A}\nu_d.
\end{equation}
The differential-reluctivity construction is therefore exact for this single-tube case, where it coincides with the Jacobian of \cite{Boomiraja2020Convergence}.

Branch $i$ of loop $x$ consists of two half-branches, $h_1$ and $h_2$, one in each cell it joins. Both carry the branch flux $\varPhi_x-\varPhi_i$. The branch MMF drop and correction reluctance are the sums of their half-branch values. Applying the chain rule gives
\begin{equation}
\label{eq:jacobian_branch_chain_rule}
\begin{aligned}
\frac{\partial\mathcal F_i}{\partial\varPhi_j}
&=\sum_{h\in\{h_1,h_2\}}\sum_k\frac{\partial\mathcal F_h}{\partial\phi_k}\frac{\partial\phi_k}{\partial\varPhi_j}\\
&\approx\widetilde{\mathcal R}_i\frac{\partial(\varPhi_x-\varPhi_i)}{\partial\varPhi_j}.
\end{aligned}
\end{equation}
The source MMF is fixed, so differentiating the loop residual gives
\begin{equation}
\label{eq:jacobian_approx_entry}
\widetilde J_{xj}=\sum_{i=1}^{4}\widetilde{\mathcal R}_i\frac{\partial(\varPhi_x-\varPhi_i)}{\partial\varPhi_j}.
\end{equation}
For the loop in Fig.~\ref{2D Mesh loop}, these entries are
\begin{equation}
\label{eq:jacobian_loop_entries}
\begin{aligned}
\widetilde J_{xx}&=\sum_{i=1}^{4}\widetilde{\mathcal R}_i,\\
\widetilde J_{xi}&=-\widetilde{\mathcal R}_i,\qquad i=1,\ldots,4.
\end{aligned}
\end{equation}
(Entries for nonadjacent loops are zero.) Thus, the approximate Jacobian is the same as the reluctance matrix in (\ref{residual eq2}), except that it is calculated using differential reluctivities, rather than apparent reluctivities. This construction is exact for linear materials, for which differential and apparent reluctivities are equal.

The approximate Jacobian is symmetric because each shared branch has the same correction reluctance in both loop equations. Its quadratic form is
\begin{equation}
\label{eq:jacobian_positive_definite}
z^T\widetilde Jz=\sum_{\text{branches}}\widetilde{\mathcal R}\,(\Delta z)^2>0,
\qquad z\ne0,
\end{equation}
where $\Delta z$ is the difference of adjacent loop values, or the loop value at a model boundary. The sum is positive because every correction reluctance is positive, and it is zero only when every loop value is zero. The approximate Jacobian is therefore symmetric positive definite and can be solved using Cholesky factorization (Subsection~\ref{subsec:linear_solver}).

Near a solution, Newton--Raphson with an exact Jacobian converges quadratically, whereas fixed-point iteration converges linearly~\cite{Chapra2015Numerical}. The inexact Newton iteration converges linearly when the error of each linearized solve stays below a fixed fraction of the residual, smaller than one, near the solution, and superlinearly when that fraction tends to zero \cite{Dembo1982Inexact}.

\subsection{Linear-System Solution Methods}
\label{subsec:linear_solver}
Each Newton--Raphson iteration solves the linear system

\begin{equation}
\label{linear system eq}
Ax=b,
\end{equation}

\noindent where $A$ is the Jacobian matrix of the iteration, $x$ is the correction, and $b$ is the negative of the residual. The matrix is stored in sparse form. The methods that use symmetry assume that $A$ is symmetric positive definite, which the approximate Jacobian is (Subsection~\ref{subsec:approx_jacobian}).

\subsubsection{Explicit Inversion}
\label{subsubsec:dense_inversion}
The most direct method is to form the inverse of $A$ and multiply, $x=A^{-1}b$, where $A^{-1}$ is found by solving $A$ for each column of the identity matrix. The inverse of a sparse matrix is generally dense, so for an $N\times N$ matrix this stores $N^2$ numbers and takes a number of operations proportional to $N^3$, even though $A$ has only a few nonzero entries per row \cite{Tinney1967Direct}. This cost is paid at every iteration. The peak memory therefore grows as 

\begin{equation}
\label{memory for explicit inversion}
M(N)=M_0+cN^2,
\end{equation}

\noindent where $M_0$ is the baseline process memory and $c$ covers the inverse and its work arrays. Factorization methods avoid forming the inverse: they split $A$ into triangular factors once, and then obtain $x$ with two substitutions.

\subsubsection{LU Factorization}
\label{subsubsec:lu_factorization}
LU factorization \cite{TURING1948ROUNDING} writes a nonsingular matrix as $PAQ=LU$, with permutation matrices $P$ and $Q$, a lower triangular matrix $L$, and an upper triangular matrix $U$, and obtains the correction by forward and backward substitution.

\subsubsection{Cholesky Factorization}
\label{subsubsec:cholesky_factorization}
Cholesky factorization uses the symmetric positive-definite structure of $A$ and writes

\begin{equation}
\label{cholesky eq}
A=LL^{\mathsf T},
\end{equation}

\noindent where $L$ is lower triangular with a positive diagonal. The correction follows from solving $Ly=b$ by forward substitution and $L^{\mathsf T}x=y$ by backward substitution. Sparse Cholesky factorization is computed here with CHOLMOD \cite{Chen2008Algorithm}, which groups columns with the same structure into supernodes. Sparse LU factorization does not exploit matrix symmetry, whereas Cholesky factorization does.

\subsubsection{Conjugate Gradient and Incomplete Cholesky Preconditioning}
\label{subsubsec:cg_iccg}
CG \cite{Hestenes1952Methods} also uses the symmetric positive-definite structure but computes the correction iteratively. Each iteration forms one product with $A$, so no factor is formed. It stops when $\lVert b-Ax_j\rVert/\lVert b\rVert$ falls below a tolerance. ICCG applies CG with a preconditioner $M=\tilde L\tilde L^{\mathsf T}\approx A$, where $\tilde L$ is an incomplete Cholesky factor \cite{Meijerink1977An}. Each ICCG iteration applies $M^{-1}$ by two triangular solves. IC(0) keeps only the entries of $\tilde L$ at positions where the lower triangle of $A$ is nonzero. The threshold variant allows fill but discards entries smaller than a drop tolerance \cite{Scott2014On}. In this paper, ICCG refers to the threshold variant. The order of the unknowns determines which entries of the incomplete factor are discarded.

\subsubsection{Reordering and Scaling}
\label{subsubsec:reordering_scaling}
The factors of a sparse matrix can have nonzero entries at positions where the matrix has none, called fill-in. Two steps applied to $A$ before factorization are therefore tested in this paper: reordering and scaling. Reordering permutes the unknowns, so that $PAP^{\mathsf T}$ is factored instead of $A$, and the elimination order determines how much fill-in appears. Minimum-degree reordering eliminates the unknown with the fewest connections first \cite{Tinney1967Direct}. Approximate minimum-degree reordering \cite{Amestoy1996An,Davis2004Algorithm} estimates that count at lower cost. Scaling replaces $A$ by $D^{-1/2}AD^{-1/2}$, where $D$ is the diagonal of $A$, giving a unit diagonal \cite{Sluis1969Condition}. It changes the entries of the matrix but not its sparsity pattern. Cholesky factorization and ICCG use approximate minimum-degree reordering, and ICCG also uses Jacobi scaling. CG and IC(0) use neither, and sparse LU factorization is computed by MATLAB's own routine. Subsection~\ref{subsec:linear_solver_results} tests these choices.

\subsection{Newton--Raphson Step Control}
\label{subsec:step_control}
The differential relative permeability falls by more than 99.9\% from its peak to one in saturated iron, so the Jacobian of a cell changes strongly with flux density. Step control addresses this by changing how the correction updates the mesh fluxes. The step length $\alpha_k$ in (\ref{Newton Raphson eq update}) is set by full-step Newton--Raphson or line search. Anderson mixing controls the step differently, by combining the current and previous Newton--Raphson results to determine the next mesh-flux state. In full-step Newton--Raphson, $\alpha_k=1$, so the full correction is applied at every iteration. In line search, the trial correction starts at $\alpha_k=1$. A trial is accepted if it reduces the normalized residual by at least the fraction $10^{-4}\alpha_k$ of its value before the step. Otherwise $\alpha_k$ is halved and the trial is repeated, down to a smallest step of $1/1024$. If no trial is accepted, the solution stops and the design is counted as not converged. This rule of halving the step until a sufficient decrease is obtained appears in Armijo's modified steepest descent algorithm for function minimization \cite{Armijo1966Minimization}.

In Anderson mixing, also called Anderson acceleration \cite{Walker2011Anderson}, each iteration still computes the Newton--Raphson correction $\Delta u_k$ of (\ref{Newton Raphson eq}), but the next state is a combination of the current result and up to 10 previous results. The combination weights sum to one and are chosen by least squares to make the combined update as small as possible. The mixed state is accepted only if its normalized residual is lower than that before the step. Otherwise, the plain Newton--Raphson update is applied and the stored history restarts from that update. All step-control methods use the same convergence criterion, initialization, and iteration limit.

\section{Evaluation}
\label{sec:evaluation}

\subsection{Study Designs}
\label{subsec:study_designs}
Three diverse magnetic gear designs, Base Design 1 (BD1), Base Design 2 (BD2), and Base Design 3 (BD3), are used for the initial analyses (Table~\ref{base designs tab}). They use M250 electrical steel for the ferromagnetic components and NdFeB N42 for the PMs. Their MECs have 245,280, 139,080, and 686,400 cells, respectively. $T_{Mods}$ denotes the total radial thickness of Rotor 2, including the bridge, so the toothed modulator portion has thickness $T_{Mods}-T_{Brg}$.

\begin{table}[!t]
\caption{Base-design specifications.}
\label{base designs tab}
\setlength\tabcolsep{3pt}
\begin{tabular}{l c c c c c}
\toprule
\textbf{Parameter} & \textbf{Description} & \textbf{BD1} & \textbf{BD2} & \textbf{BD3} & \textbf{Units}\\
\midrule
$P_{1}$ & \multicolumn{1}{l}{Rotor 1 pole pairs} & 11 & 4 & 6 &   \\
$P_{3}$ & \multicolumn{1}{l}{Rotor 3 pole pairs} & 45 & 34 & 98 &   \\
$r_{out}$ & \multicolumn{1}{l}{Rotor 3 back iron outer radius} & 150 & 175 & 200 & mm \\
$L$ & \multicolumn{1}{l}{Axial stack length} & 1 & 1 & 1 & m \\
$T_{BI1}$ & \multicolumn{1}{l}{Rotor 1 back iron thickness}  & 20 & 35 & 40 & mm  \\
$T_{PM1}$ & \multicolumn{1}{l}{Rotor 1 PM thickness}  & 9 & 5 & 13 & mm \\
$T_{AG1}$ & \multicolumn{1}{l}{Inner air gap thickness}  & 0.5 & 2 & 1 & mm \\
$T_{Mods}$ & \multicolumn{1}{l}{Modulator thickness$^{*}$}  & 11 & 17 & 14 & mm \\
$T_{Brg}$ &  \multicolumn{1}{l}{Bridge thickness}  & 0.5 & 1 & 1.5 & mm \\
$T_{AG2}$ & \multicolumn{1}{l}{Outer air gap thickness}  & 0.5 & 2 & 1 & mm \\
$T_{PM3}$ & \multicolumn{1}{l}{Rotor 3 PM thickness}  & 7 & 5 & 7 & mm  \\
$T_{BI3}$ & \multicolumn{1}{l}{Rotor 3 back iron thickness}  & 20 & 30 & 25 & mm \\
\bottomrule
\multicolumn{5}{@{}l@{}}{\scriptsize * Bridge included.} \\
\end{tabular}
\end{table}

To evaluate the nonlinear MEC model across the design space, a population of 1,024 bridged RFMG designs is drawn from the ranges in Table~\ref{Parameter sweep ranges}. The active outer radius is fixed, so all designs share one envelope and their torques are directly comparable. The magnets fill the full pole pitch, which gives the largest magnet flux and the most severe saturation.

\begin{table}[!ht]
\setlength{\tabcolsep}{4pt}
\caption{Parameter ranges.}
\centering
\begin{tabularx}{\columnwidth}{@{}l X c c c@{}}
\toprule
\textbf{Parameter} & \textbf{Description} & \textbf{Range} & \textbf{Resolution} & \textbf{Units} \\
\midrule
$G_{r}$ & Integer part of gear ratio & 4--16 & 1 & \\
$P_{1}$ & Rotor 1 pole pairs & 3--18 & 1 & \\
$T_{BI1}$ & Rotor 1 back iron thickness & 5--40 & 0.1 & mm \\
$T_{PM1}$ & Rotor 1 PM thickness & 3--13 & 0.1 & mm \\
$T_{AG}$ & Air gap thickness, each gap & 0.5--2 & 0.05 & mm \\
$T_{Mods}$ & Modulator thickness$^{*}$ & 8--17 & 0.1 & mm \\
$T_{Brg}$ & Bridge thickness & 0.5--2 & 0.05 & mm \\
$\alpha_{Mods}$ & Modulator fill factor & 0.3--0.7 & 0.01 & \\
$T_{PM3}$ & Rotor 3 PM thickness & 1.5--15 & 0.1 & mm \\
$T_{BI3}$ & Rotor 3 back iron thickness & 10--30 & 0.1 & mm \\
$P_{3}$ & Rotor 3 pole pairs, from (\ref{P_LS eq}) & $\leq 130$ & & \\
$r_{out}$ & Rotor 3 back iron outer radius & 150 & & mm \\
$L$ & Axial stack length & 1 & & m \\
$\alpha_{PM}$ & PM fill factor & 1 & & \\
\bottomrule
\multicolumn{5}{@{}l@{}}{\scriptsize * Bridge included.} \\
\end{tabularx}
\label{Parameter sweep ranges}
\end{table}

The modulator count for an RFMG is $Q_2=P_{1}+P_{3}$, where $P_{1}$ and $P_{3}$ are the PM pole-pair counts of Rotors 1 and 3 \cite{atallah2001}, which gives the harmonic matching required for synchronous torque transmission. With the modulators (Rotor 2) stationary, the gear ratio is $-P_{3}/P_{1}$ \cite{Praslicka2022Review}, and Rotors 1 and 3 serve as the high-speed and low-speed rotors, respectively, spinning in opposite directions. As in \cite{Johnson2018Parameterizeda,Johnson2021Parameterizeda}, $G_{r}$ is the integer part of the gear ratio and relates the pole pair counts according to
\begin{equation}
\label{P_LS eq}
P_{3} = \left\{
\begin{aligned}
& G_{r}P_{1}+1, && (G_{r}+1)P_{1} \text{ odd} \\
& G_{r}P_{1}+2, && (G_{r}+1)P_{1} \text{ even}
\end{aligned}
\right.
\end{equation}
This rule gives an even modulator count to balance the net magnetic forces on each rotor and, for $P_{1}\geq3$, a noninteger gear ratio to reduce torque ripple \cite{Praslicka2022Review}. The limit on $P_{3}$ keeps each Rotor 3 PM at least about 3~mm wide at the air gap.

The designs are drawn from a scrambled quasi-random Sobol sequence \cite{Sobol1967On,Owen1998Scrambling}, which spreads them evenly across all parameter ranges in Table \ref{Parameter sweep ranges} at once, with every allowed design equally likely. The sampled ranges produce designs with and without heavy saturation.

Each design is evaluated using the nonlinear MEC and the nonlinear FEA reference of Subsection~\ref{subsec:fea_reference}, and again with its bridge removed ($T_{Brg}=0$) and all other dimensions unchanged, which gives a matched bridge-free design. All MEC analyses, including the linear MEC of Subsection~\ref{subsec:bh_material}, use the mesh of Subsection~\ref{subsec:meshing}, and all Newton--Raphson iterations use the approximate Jacobian of Subsection~\ref{subsec:approx_jacobian}.

Each MEC and FEA run was allocated two processor threads and 10~GiB of memory. The processor frequency was limited to 2~GHz. For each design, the total time is the wall-clock time of the whole run, the solve time is that of the nonlinear or linear solution, and the setup time, which covers building the geometry, mesh, and model, is the difference.

\subsection{Mesh-Flux versus Node-MMF}
\label{subsec:formulation_comparison}
Both formulations describe the same MEC, so the comparison selects the one whose iteration converges. The node-MMF iteration, which uses the permeabilities from the previous iteration, and the mesh-flux iteration were each run with the same settings, full Newton--Raphson steps for 50 iterations. After 50 iterations, the node-MMF normalized residuals were between $8.3\times10^{2}$ and $1.4\times10^{3}$. The mesh-flux normalized residuals were between $9.8\times10^{-5}$ and $5.5\times10^{-4}$ (Table~\ref{fixed_point_comparison_tab}).

Given these normalized residuals, the mesh-flux formulation is used throughout this paper. The circular dependency between branch flux and cell permeability likely explains the large node-MMF normalized residual (Subsection~\ref{subsec:node_mmf}).

\subsection{Inexact Newton Iteration versus Fixed-Point Iteration}
\label{subsec:newton_vs_fixed_point}
To assess whether Newton--Raphson improves convergence despite using an approximate Jacobian, it was compared with fixed-point iteration (Subsection~\ref{subsec:newton_raphson}) with and without relaxation on the base designs. After 50 iterations, full-step Newton--Raphson reached much lower normalized residuals on all three designs than every fixed-point variant evaluated (Table~\ref{fixed_point_comparison_tab}).

\begin{table}[!t]
\centering
\footnotesize
\setlength{\tabcolsep}{3pt}
\caption{Final normalized residuals of full-step Newton--Raphson and fixed-point iteration for BD1--BD3 after at most 50 iterations. Each linear solve uses sparse Cholesky factorization.}
\label{fixed_point_comparison_tab}
\begin{tabular}{llrrr}
\toprule
Method & $\beta$ & BD1 & BD2 & BD3 \\
\midrule
Newton--Raphson & --- & $9.8\times10^{-5}$ & $5.5\times10^{-4}$ & $3.6\times10^{-4}$ \\
\midrule
Fixed point & 1 & 7.3 & 13 & 7.2 \\
 & 0.5 & 0.49 & 0.62 & 0.79 \\
 & 0.25 & 0.045 & 0.076 & 0.059 \\
 & 0.1 & $1.4\times10^{-3}$ & $4.3\times10^{-3}$ & $2.8\times10^{-3}$ \\
 & 0.05 & 0.11 & 0.22 & 0.063 \\
\bottomrule
\end{tabular}
\end{table}

\subsection{Linear Solution Methods}
\label{subsec:linear_solver_results}

All linear-solver tests ran 50 full-step Newton--Raphson iterations on BD1--BD3 from the same initial state. Reordering reduced the solve time and peak memory of both sparse Cholesky factorization and ICCG, especially for BD3, whose plain Cholesky run exceeded the 10~GiB memory limit (Table~\ref{reordering_ablation_tab}). Jacobi scaling did not change the time or memory of either solver. Reordering slowed IC(0) (Table~\ref{ic_settings_tab}), consistent with the known effect of ordering on incomplete factorizations \cite{Duff1989Effect}. Reordering is therefore used for sparse Cholesky factorization and ICCG throughout.

\begin{table}[!t]
\caption{Solve time and peak memory of sparse Cholesky factorization and ICCG with and without reordering for BD1--BD3.}
\label{reordering_ablation_tab}
\centering
\footnotesize
\setlength{\tabcolsep}{4pt}
\begin{tabular}{llrrrr}
\toprule
 & & \multicolumn{2}{c}{Time (s)} & \multicolumn{2}{c}{Memory (GiB)} \\
\cmidrule(lr){3-4}\cmidrule(l){5-6}
Solver & Design & Plain & Reordered & Plain & Reordered \\
\midrule
\multirow[c]{3}{*}{Cholesky} & BD1 & 3306 & 109 & 8.37 & 1.71 \\
 & BD2 & 1039 & 67 & 4.00 & 1.45 \\
 & BD3 & 23,840 & 248 & 37.86 & 3.01 \\
\midrule
\multirow[c]{3}{*}{ICCG} & BD1 & 295 & 154 & 1.82 & 1.69 \\
 & BD2 & 164 & 86 & 1.55 & 1.48 \\
 & BD3 & 1485 & 489 & 3.29 & 2.91 \\
\bottomrule
\end{tabular}
\end{table}

\begin{table}[!t]
\caption{Solve times of ICCG settings for BD1--BD3.}
\label{ic_settings_tab}
\centering
\footnotesize
\setlength{\tabcolsep}{4pt}
\begin{tabular}{llrrr}
\toprule
ICCG setting & Scaling and reordering & BD1 (s) & BD2 (s) & BD3 (s) \\
\midrule
IC(0) & No & 958 & 689 & 6374 \\
IC(0) & Yes & 1586 & 1131 & 6544 \\
Threshold $10^{-3}$ & Yes & 437 & 232 & 2781 \\
Threshold $10^{-4}$ & Yes & 214 & 96 & 943 \\
Threshold $10^{-5}$ & Yes & 153.9 & 84.1 & 472.2 \\
\bottomrule
\end{tabular}
\end{table}

\subsubsection{Explicit Inversion versus Sparse Cholesky Factorization}
\label{subsubsec:dense_vs_cholesky}

\begin{figure}[!t]
\centering
\includegraphics{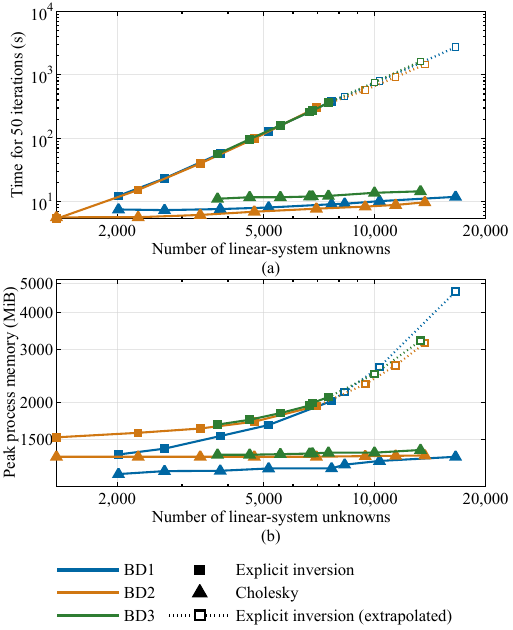}
\caption{Explicit inversion versus sparse Cholesky factorization on coarsened meshes of BD1--BD3. (a) Total time and (b) peak process memory. Hollow squares and dotted lines are estimates for levels not run with explicit inversion.}
\label{dense_inverse_cholesky_fig}
\end{figure}

At 8 bytes per number, the inverse alone occupies about 671~GiB for 300,000 unknowns, less than half the number of unknowns for BD3. Explicit inversion is therefore compared with sparse Cholesky factorization on coarsened BD1--BD3 meshes. Explicit inversion was run up to about 8,000 unknowns, where 50 iterations took over 300~s, and sparse Cholesky factorization ran at every level. For larger meshes, the run time and peak memory of explicit inversion are estimated by fitting a power law and (\ref{memory for explicit inversion}) to the executed levels of each base design. The speedup of sparse Cholesky factorization grew with mesh size, up to 42$\times$ at 7,636 unknowns (Fig.~\ref{dense_inverse_cholesky_fig}(a)). Its peak process memory was up to 25\% lower (Fig.~\ref{dense_inverse_cholesky_fig}(b)).

\subsubsection{Sparse Cholesky Factorization versus Alternative Linear Solution Methods}
\label{subsubsec:cholesky_vs_alternatives}

Sparse Cholesky factorization, CG, ICCG, and sparse LU were compared for BD1--BD3. Among the ICCG settings reported in Table~\ref{ic_settings_tab}, a drop tolerance of $10^{-5}$ gave the shortest solve time for every design and was selected for comparison with the other solvers. Table~\ref{linear_solver_comparison_tab} reports the solve time, final normalized residual, and Rotor 1 torque of each combination. Sparse Cholesky factorization was the fastest method in every comparison. Sparse LU factorization took 1.3$\times$ to 2.2$\times$ as long, CG 7.9$\times$ to 8.8$\times$, and ICCG 1.2$\times$ to 1.9$\times$.

CG was also less reliable. It did not reach its inner tolerance of $10^{-8}$ in at least one update of every case. For BD3, its iterations ended far from the convergence criterion (Table~\ref{linear_solver_comparison_tab}), and its final Rotor 1 torque differed from that of the other methods by 62\%.

\begin{table}[!t]
\caption{Linear-solver comparison for BD1--BD3. ICCG uses a drop tolerance of $10^{-5}$. Torque is for Rotor 1.}
\label{linear_solver_comparison_tab}
\centering
\footnotesize
\renewcommand{\arraystretch}{1.05}
\setlength{\tabcolsep}{3.2pt}
\begin{tabular}{llrrr}
\toprule
Design & Method & Solve time (s) & \makecell{Final normalized\\residual} & Torque (Nm) \\
\midrule
\multirow[c]{4}{*}{BD1} & Cholesky & 108.6 & $1.12\times10^{-4}$ & 2594.2 \\
 & CG & 860 & $3.46\times10^{-4}$ & 2593.9 \\
 & ICCG & 153.9 & $1.13\times10^{-4}$ & 2594.2 \\
 & LU & 169 & $1.11\times10^{-4}$ & 2594.2 \\
\midrule
\multirow[c]{4}{*}{BD2} & Cholesky & 67.3 & $6.69\times10^{-4}$ & 714.2 \\
 & CG & 545 & $9.78\times10^{-4}$ & 714.0 \\
 & ICCG & 84.1 & $6.75\times10^{-4}$ & 714.2 \\
 & LU & 87.8 & $6.53\times10^{-4}$ & 714.2 \\
\midrule
\multirow[c]{4}{*}{BD3} & Cholesky & 248.2 & $4.76\times10^{-4}$ & 887.5 \\
 & CG & 2187 & $2.49\times10^{-2}$ & 338.5 \\
 & ICCG & 472.2 & $4.82\times10^{-4}$ & 887.5 \\
 & LU & 558 & $4.78\times10^{-4}$ & 887.5 \\
\bottomrule
\end{tabular}
\end{table}

\subsection{Nonlinear Convergence}
\label{subsec:convergence}
Full-step Newton--Raphson, line search, and Anderson mixing are compared to assess convergence and computational cost. All three use sparse Cholesky factorization for each linearized solve.

\begin{figure*}[!t]
\centering
\includegraphics[width=\textwidth]{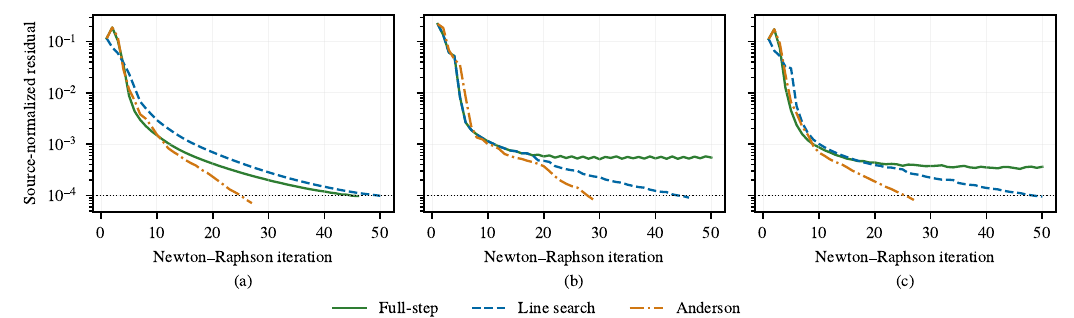}
\caption{Normalized residual histories of (a) BD1, (b) BD2, and (c) BD3. The dotted line marks the $10^{-4}$ convergence criterion.}
\label{base_design_residual_histories_fig}
\end{figure*}

Fig.~\ref{base_design_residual_histories_fig} shows the normalized residual histories of the three base designs. With full steps, the normalized residuals of BD2 and BD3 stop decreasing and oscillate without converging. Line search removes this stagnation and converges BD2 and BD3. When full steps already reduce the normalized residual, line search accepts every full step and the two histories coincide, as in the first 18 iterations for BD2. For BD1, however, it reduces the normalized residual more slowly than full steps and does not quite converge within 50 iterations. Anderson mixing converges every base design in the fewest iterations, 27 to 29.

\begin{figure}[!t]
\centering
\includegraphics{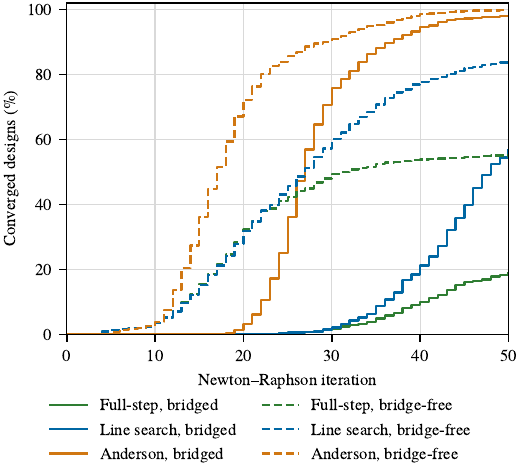}
\caption{Percentage of the 1,024 bridged and 1,024 bridge-free designs converged by each iteration.}
\label{newton_convergence_outcome_fig}
\end{figure}

Fig.~\ref{newton_convergence_outcome_fig} shows the percentage of the 1,024 bridged and 1,024 bridge-free designs of Subsection~\ref{subsec:study_designs} that have reached the convergence criterion by each iteration, and Table~\ref{newton_method_population_summary_tab} summarizes the final counts, iteration statistics, and update cost. With full steps, most bridged designs do not converge within 50 iterations. Line search converges more designs, and for the designs that all three methods converge it needs the same number of iterations as full steps. Anderson mixing converges nearly every design in both families and needs the fewest iterations. Although Anderson mixing increases the cost per iteration, it reduces the total solution time by reducing the number of iterations.

\begin{table*}[!t]
\caption{Nonlinear solution methods for 1,024 bridged and 1,024 bridge-free designs. Iteration-count and solve-time statistics use the 186 bridged and 552 bridge-free designs converged by all three methods. Solve-time ratios compare mean solve times for each family. Iteration-time ratios compare mean times for matched design--iteration pairs across all three methods, pooled across both families regardless of final convergence status. All ratios are normalized to full-step Newton--Raphson.}
\label{newton_method_population_summary_tab}
\centering
\footnotesize
\setlength{\tabcolsep}{2pt}
\begin{tabular}{llccccccc}
\toprule
\multirow{2}{*}{Method} & \multirow{2}{*}{Family} & \multirow{2}{*}{Converged} & \multicolumn{2}{c}{Iterations} & \multicolumn{2}{c}{Solve time (s)} & \multirow{2}{*}{\makecell{Solve\\time ratio}} & \multirow{2}{*}{\makecell{Iteration\\time ratio}} \\
\cmidrule(lr){4-5}\cmidrule(lr){6-7}
 & & & Median & 95th & Mean & 95th & & \\
\midrule
\multirow{2}{*}{Full-step} & Bridged & 193 & 40 & 48 & 87.5 & 172 & 1.000$\times$ & \multirow{2}{*}{1.000$\times$} \\
 & Bridge-free & 564 & 19 & 35 & 43.0 & 91 & 1.000$\times$ & \\
\midrule
\multirow{2}{*}{Line search} & Bridged & 580 & 40 & 48 & 90.5 & 183 & 1.034$\times$ & \multirow{2}{*}{1.004$\times$} \\
 & Bridge-free & 861 & 19 & 35 & 41.8 & 88 & 0.971$\times$ & \\
\midrule
\multirow{2}{*}{Anderson} & Bridged & 1,003 & 23.5 & 26 & 67.6 & 134 & 0.773$\times$ & \multirow{2}{*}{1.215$\times$} \\
 & Bridge-free & 1,022 & 15 & 20 & 41.5 & 90 & 0.965$\times$ & \\
\bottomrule
\end{tabular}
\end{table*}

Bridged designs are consistently harder to converge than bridge-free designs. With every method, fewer bridged designs reach the criterion, and those that do need more iterations (Table~\ref{newton_method_population_summary_tab}). Subsection~\ref{subsec:bridge_effects} relates this to the higher peak iron flux density and larger saturated iron fraction of the bridged designs.

\subsection{FEA Reference}
\label{subsec:fea_reference}
The FEA reference is calculated with two-dimensional nonlinear FEA in NGSolve \cite{Schoberl2014NGSolve}, using second-order finite elements on a mesh that Netgen \cite{Schoberl1997NETGEN} generates from each design's geometry. The nonlinear equations are solved using Newton--Raphson iteration with line search, under the same convergence criterion as the MEC (Subsection~\ref{subsec:convergence_criterion}). The mesh remains fixed during the nonlinear solution, so each design requires one nonlinear FEA solution, matching the one MEC solution, and the FEA time depends only on the mesh resolution, selected below by a convergence study. Commercial packages such as ANSYS Maxwell instead refine the mesh adaptively, solving repeatedly until an error target is met. Consequently, speedups measured against them are inflated by these repeated solutions \cite{Yamazaki2003Adaptive,Johnson2018Parameterizeda}.

\begin{figure*}[!t]
\centering
\setlength{\abovecaptionskip}{2pt}
\includegraphics[width=\textwidth]{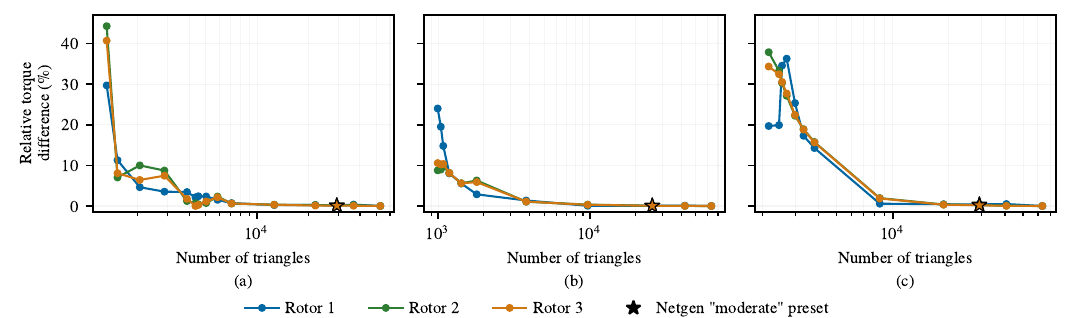}
\caption{Torque convergence for (a) BD1, (b) BD2, and (c) BD3, using fixed Netgen meshes and second-order finite elements, relative to each design's highest-resolution mesh. Stars mark the preset used for the FEA reference.}
\label{fea_mesh_convergence_fig}
\end{figure*}

FEA mesh sensitivity is evaluated for BD1--BD3 by varying the resolution. Fig.~\ref{fea_mesh_convergence_fig} shows the absolute relative torque difference against triangle count, with the highest-resolution mesh of each design as the baseline. The mesh used for the FEA reference differs from that baseline by less than 0.4\% for all three torque outputs in each design.

\subsection{Comparisons with FEA for Base Designs}
\label{subsec:base_design_fea}

\begin{figure*}[!t]
\centering
\includegraphics[width=\textwidth]{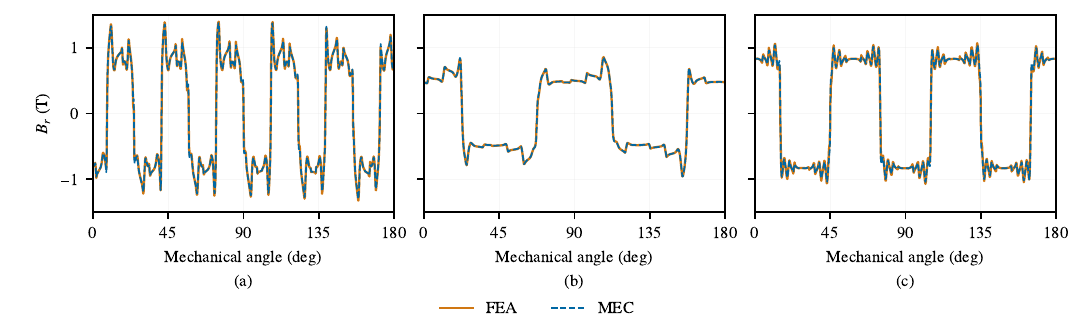}
\caption{Radial flux density from nonlinear FEA and MEC on a half circle at the radial middle of the inner air gap for (a) BD1, (b) BD2, and (c) BD3.}
\label{radial_flux_comparison}
\end{figure*}

Fig.~\ref{radial_flux_comparison} compares the radial flux-density distributions obtained from nonlinear FEA and the nonlinear MEC along a half circle at the radial middle of the inner air gaps of the three base designs. Table~\ref{air_gap_field_comparison_tab} compares the magnitudes of the $P_1$ and $P_3$ harmonics of the radial and tangential flux density around the full circle in both air gaps. These are the primary torque-producing harmonics. The points are the angular cell centers of the MEC mesh, so the harmonics are evaluated over the actual angles. The two models agree closely. Every harmonic above 0.1~T differs in magnitude by at most 3.3\% and in phase by at most 0.8\textdegree{}. The largest relative difference, 11.2\%, occurs in a $P_3$ harmonic of only 0.008~T in the inner air gap of BD3.

\begin{table}[!t]
\centering
\footnotesize
\setlength{\tabcolsep}{2.5pt}
\caption{$P_1$ and $P_3$ harmonics of radial and tangential flux density for a circle at the radial middle of each air gap. Magnitudes are in tesla. Diff. is the percent difference in magnitude relative to FEA.}
\label{air_gap_field_comparison_tab}
\begin{tabular}{@{}lllrrrrrr@{}}
\toprule
 & & & \multicolumn{3}{c}{$P_1$ harmonic} & \multicolumn{3}{c}{$P_3$ harmonic} \\
\cmidrule(lr){4-6}\cmidrule(l){7-9}
Design & Air gap & Field & FEA & MEC & Diff. (\%) & FEA & MEC & Diff. (\%) \\
\midrule
 \multirow{4}{*}{BD1} & \multirow{2}{*}{Inner} & $B_r$ & 1.098 & 1.098 & 0.0 & 0.045 & 0.044 & $-$3.6 \\
  &  & $B_\theta$ & 0.323 & 0.321 & $-$0.4 & 0.044 & 0.044 & $-$1.0 \\
  & \multirow{2}{*}{Outer} & $B_r$ & 0.354 & 0.355 & 0.2 & 1.123 & 1.124 & 0.1 \\
  &  & $B_\theta$ & 0.193 & 0.193 & 0.3 & 0.457 & 0.450 & $-$1.5 \\
\midrule
 \multirow{4}{*}{BD2} & \multirow{2}{*}{Inner} & $B_r$ & 0.669 & 0.668 & $-$0.1 & 0.018 & 0.018 & $-$1.8 \\
  &  & $B_\theta$ & 0.133 & 0.134 & 0.6 & 0.016 & 0.017 & 0.7 \\
  & \multirow{2}{*}{Outer} & $B_r$ & 0.406 & 0.406 & 0.0 & 0.833 & 0.836 & 0.4 \\
  &  & $B_\theta$ & 0.067 & 0.066 & $-$1.1 & 0.394 & 0.400 & 1.5 \\
\midrule
 \multirow{4}{*}{BD3} & \multirow{2}{*}{Inner} & $B_r$ & 1.059 & 1.059 & 0.0 & 0.008 & 0.007 & $-$11.2 \\
  &  & $B_\theta$ & 0.226 & 0.226 & 0.1 & 0.008 & 0.008 & 1.9 \\
  & \multirow{2}{*}{Outer} & $B_r$ & 0.627 & 0.627 & 0.1 & 0.827 & 0.831 & 0.5 \\
  &  & $B_\theta$ & 0.157 & 0.157 & $-$0.5 & 0.473 & 0.458 & $-$3.3 \\
\bottomrule
\end{tabular}
\end{table}

\subsection{Population Accuracy and Cost}
\label{subsec:population_accuracy}
\begin{figure}[!t]
\centering
\includegraphics{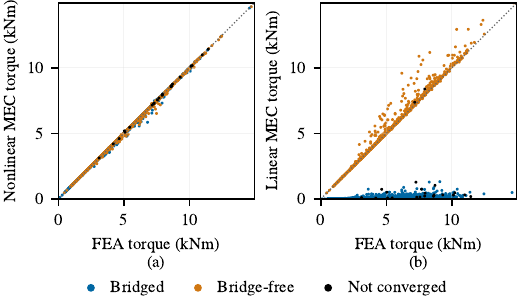}
\caption{(a) Nonlinear MEC and (b) linear MEC versus FEA Rotor 3 torque for all designs. The dotted line marks an ideal match between MEC and FEA.}
\label{MEC_FEA_ranking}
\end{figure}

Each nonlinear MEC result obtained with Anderson mixing is compared with the FEA result for the same design. The comparison uses all 1,024 bridged and 1,024 bridge-free designs. The torque discrepancy of each design is $|\tau_{\mathrm{MEC}}-\tau_{\mathrm{FEA}}|/|\tau_{\mathrm{FEA}}|$, evaluated separately for the Rotor 1, Rotor 2, and Rotor 3 torques. Fig.~\ref{MEC_FEA_ranking} compares the MEC and FEA Rotor 3 torque predictions, Table~\ref{reference FEA comparison} summarizes the discrepancies, and Table~\ref{population_cost_tab} gives the computational cost.

FEA met the convergence criterion for every design, and the MEC met it for 1,003 bridged and 1,022 bridge-free designs. The other 21 bridged and 2 bridge-free designs reached the 50-iteration limit before meeting the criterion, with final normalized residuals of up to $5.0\times10^{-4}$. Their median and largest Rotor~3 torque discrepancies were 0.19\% and 3.0\%, which is well within the range of the converged designs. In a separate check with a limit of 100 iterations, all but one of them met the convergence criterion, and no rotor torque changed by more than 0.004\% from its value at 50 iterations. Not meeting the criterion therefore indicates lower confidence in these solutions rather than failure. These designs are marked separately in Fig.~\ref{MEC_FEA_ranking}.

The nonlinear MEC and FEA agree well throughout the range of Rotor 3 torques (Fig.~\ref{MEC_FEA_ranking}), and the two families have nearly the same discrepancy distributions (Table~\ref{reference FEA comparison}). The largest Rotor 3 torque discrepancies occur in designs with the smallest Rotor 1 pole-pair count (the 20 cases with the largest discrepancy have $P_1=3$) and a thin Rotor 1 back iron. The low pole-pair count produces large pole arcs, and the thin back irons are prone to saturation \cite{Johnson2018Parameterizeda}, which lowers their permeability, raises the reluctance of the flux return path, and reduces the air-gap flux and torque \cite{Lee2021Analytic}. Because the Rotor 1 back iron is thin and heavily saturated in these designs, more of the Rotor 1 PM flux leaks into the bore than in most designs. Thus, the two radial layers used in the bore (Subsection~\ref{subsec:meshing}) may not have captured this leakage accurately, which would increase the discrepancy. However, these designs are a small fraction of the population and are not near the maximum torques predicted by either model (Fig.~\ref{MEC_FEA_ranking}).

Beyond predicting the torque of each design, the MEC should also order designs correctly by torque for design studies and optimization. Agreement between the MEC and FEA models' rankings is measured by the Spearman rank correlation \cite{Spearman1904Proof}. The correlation is at least 0.9996 for every torque output in both families. For each rotor torque, both models identify the same highest-torque design in each family.

The linear MEC of Subsection~\ref{subsec:bh_material} is also compared with FEA for the same designs. For the bridge-free designs its median discrepancy is close to that of the nonlinear MEC, but its 95th percentile Rotor 3 torque discrepancy is 13.0\%. For the bridged designs it predicts only about 1\% of the FEA torque for every torque. Bridges form flux leakage paths \cite{Li2019Designing}, and with unsaturated iron they short-circuit the permanent magnet flux, so almost no flux reaches the air gaps. The linear model is therefore adequate for most bridge-free designs but not for strongly saturated ones, and completely unusable for bridged designs, whereas the nonlinear MEC has nearly the same accuracy in both families (Table~\ref{reference FEA comparison}).

The nonlinear MEC runs about twice as fast as the FEA (Table~\ref{population_cost_tab}). Setup takes a similar amount of time in every model, so the speedup primarily comes from solving the equations faster. This speedup is smaller than the 7.5 to 20$\times$ reported for other nonlinear MECs \cite{Amrhein20093,Korman2025Distributed}. Those speedups were measured against FEA with different meshes and settings, so they cannot be compared directly with the speedup here (Section~\ref{sec:introduction}).

\begin{table}[!t]
\centering
\footnotesize
\setlength{\tabcolsep}{3pt}
\caption{MEC torque discrepancy relative to FEA for all designs.}
\label{reference FEA comparison}
\begin{tabular}{llrrrrrr}
\toprule
& & \multicolumn{3}{c}{Nonlinear MEC (\%)} & \multicolumn{3}{c}{Linear MEC (\%)} \\
\cmidrule(lr){3-5}\cmidrule(lr){6-8}
Family & Torque & Median & 95th & Max & Median & 95th & Max \\
\midrule
\multirow{3}{*}{Bridged} & Rotor 1 & 0.54 & 3.13 & 13.0 & 98.6 & 99.4 & 99.9 \\
 & Rotor 2 & 0.18 & 1.57 & 14.1 & 98.6 & 99.4 & 99.6 \\
 & Rotor 3 & 0.22 & 1.59 & 14.1 & 98.6 & 99.4 & 99.7 \\
\midrule
\multirow{3}{*}{Bridge-free} & Rotor 1 & 0.66 & 3.65 & 12.3 & 0.77 & 14.0 & 51.5 \\
 & Rotor 2 & 0.22 & 1.49 & 12.9 & 0.55 & 13.4 & 54.9 \\
 & Rotor 3 & 0.23 & 1.49 & 13.0 & 0.51 & 13.0 & 55.2 \\
\bottomrule
\end{tabular}
\end{table}

\begin{table}[!t]
\centering
\footnotesize
\setlength{\tabcolsep}{3.8pt}
\caption{Total computation times for all designs of Table~\ref{reference FEA comparison}. The nonlinear MEC times include the 23 designs whose solution did not converge within the limit of 50 iterations. Speedup is FEA total time divided by the model's total time.}
\label{population_cost_tab}
\begin{tabular}{llrrrr}
\toprule
Family & Model & Setup (h) & Solve (h) & Total (h) & Speedup \\
\midrule
\multirow{3}{*}{Bridged} & Nonlinear MEC & 3.4 & 16.5 & 19.9 & 2.17$\times$ \\
 & Linear MEC & 2.9 & 0.5 & 3.3 & 13.01$\times$ \\
 & FEA & 3.2 & 39.9 & 43.1 & --- \\
\midrule
\multirow{3}{*}{Bridge-free} & Nonlinear MEC & 3.4 & 10.5 & 13.9 & 1.84$\times$ \\
 & Linear MEC & 2.8 & 0.5 & 3.2 & 7.96$\times$ \\
 & FEA & 2.9 & 22.7 & 25.7 & --- \\
\bottomrule
\end{tabular}
\end{table}

\subsection{Effect of Modulator Bridges}
\label{subsec:bridge_effects}
\begin{figure}[!t]
\centering
\includegraphics{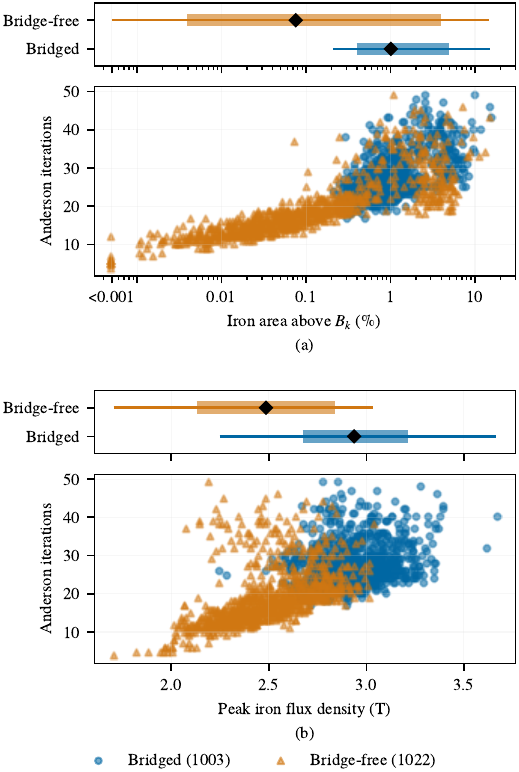}
\caption{Anderson mixing iterations versus (a) saturated iron fraction and (b) peak iron flux density for 1,003 bridged and 1,022 bridge-free converged designs. Panel (a) uses a logarithmic axis, with values below 0.001\% grouped at the labeled floor. The strips above show each family's range (thin lines), 5th--95th percentiles (thick segments), and median (diamonds).}
\label{saturation_iterations_first256}
\end{figure}

Fig.~\ref{saturation_iterations_first256} plots the Anderson mixing iteration count of the converged designs against two saturation measures, the peak iron flux density and the saturated iron fraction, which is the fraction of iron area where the flux density exceeds $B_k$ (1.95 T). Bridged designs have higher median values of both and need more iterations than bridge-free designs (Subsection~\ref{subsec:convergence}). Within each family, designs with more saturated iron also need more iterations. Between the two families, designs with similar saturated iron fractions and peak flux densities tend to require similar ranges of iterations to converge. Thus, the difference in convergence behavior is likely due to the impact of the bridge on the saturated iron fraction and peak flux density.

\section{Conclusion}

This work proposed solution methods for a parameterized nonlinear MEC: an approximate Jacobian that permits sparse Cholesky factorization, and Anderson mixing. It compared them with alternative methods on bridged RFMGs paired with the same designs without the bridge.

The exact Jacobian of the MEC cells is not symmetric. However, omitting its inter-branch terms gives an approximate Jacobian that is symmetric positive definite and reduces to the exact Jacobian for single flux tubes. The omitted terms did not cost Newton--Raphson its advantage over fixed-point iteration, which failed to converge with or without relaxation. The approximate Jacobian also made sparse Cholesky factorization applicable, which was the fastest linear solver tested. Reordering the unknowns shortened its solves and those of ICCG, especially for the design with the most unknowns to solve for. With this approximate Jacobian, step control decided reliability: full-step Newton--Raphson often failed, and Anderson mixing converged nearly every design in the fewest iterations. Bridge saturation made every method converge less often and more slowly than for the same designs without the bridge.

With these methods, the nonlinear MEC agreed closely with nonlinear FEA in torque and design ranking for both bridged and bridge-free designs, and it was about twice as fast as fixed-mesh FEA in total computation time, so it is suited to population studies. The linear MEC was adequate for most bridge-free designs but failed for bridged designs because it cannot represent bridge saturation.

\section*{Acknowledgment}
This work was partially supported under US National Science Foundation Award \#2440024.  Any opinions, findings, and conclusions or recommendations expressed in this material are those of the authors and do not necessarily reflect the views of the National Science Foundation or the US government.

The authors used ChatGPT (OpenAI) to edit the language of this manuscript. The authors reviewed all edited text and take full responsibility for the content.

\ifmanuscriptbibliographyavailable
  \bibliographystyle{IEEEtran}
  \bibliography{references}
\fi

\end{document}